\documentclass[aps,twocolumn,groupedaddress,nofootinbib,amsmath,amssymb]{revtex4}
\usepackage{dsfont}
\usepackage{bm}
\usepackage{mathrsfs}
\usepackage{xcolor}

\begin{document}

\title{From static to Vaidya solutions in scalar tensor theories}

\author{Mokhtar Hassaine, Ulises Hernandez-Vera and Franco Lara-Munoz}

\affiliation{Instituto de Matem\'atica, Universidad de Talca,
Casilla 747, Talca, Chile}

\begin{abstract}
We consider some classes of Horndeski theories in four dimensions
for which a certain combination of the Einstein equations within a
spherical ansatz splits into two distinct branches. Recently, for
these theories, some integrability and compatibility conditions have
been established which have made it possible to obtain black hole
solutions depending on a single integration constant identified as
the mass. Here, we will show that these compatibility conditions can
be generalized to accommodate a time dependence by promoting the
constant mass to an arbitrary function of the retarded (advanced)
time. As a direct consequence, we prove that all the static black
hole solutions can be naturally promoted to non static Vaidya-like
solutions. We extend this study in arbitrary higher dimensions where
the pure gravity part is now described by the Lovelock theory and,
where the scalar field action enjoyed the conformal invariance. For
these theories, the splitting in two branches is also effective, and
we show that their known static black hole solutions can as well be
promoted to Vaidya-like solutions.

\end{abstract}

\maketitle

%%%%%%%%%%%%%%%%%%%%%%%%
\section{Introduction}
%%%%%%%%%%%%%%%%%%%%%%%%

One of the important generalizations of the Schwarzschild solution
with the aim of describing the exterior of a star which is either
emitting or absorbing null dusts is undoubtedly  the so-called
Vaidya solution \cite{Vaidya}. This "radiating" solution can be
conveniently written in the Eddington-Finkelstein coordinates by
promoting the Schwarzschild constant mass $M$ to a function of the
retarded (advanced) time
$$
ds^2=-\left(1-\frac{2M(u)}{r}\right)du^2+2\epsilon
dudr+r^2d\Omega_2^2,
$$
where $\epsilon=\pm 1$ describes incoming (resp. outgoing) radiation
shells. This in turn implies that the exterior field of the
radiating star behaves as a pure radiation field with an
energy-momentum $T_{\mu\nu}$ that has component only along the
retarded (advanced) time,
\begin{eqnarray}
\label{Vaidyacond}
 T_{\mu\nu}=-\frac{2\dot{M}(u)}{r^2}k_{\mu}k_{\nu},
\end{eqnarray}
with ${\bf k}=\partial_r$, and where the dot stands for the
derivative with respect to $u$. In contrast with the Schwarzschild
metric, the Vaidya spacetime has no timelike Killing vector field,
and hence it is a non static  metric. One of the interesting aspect
of this solution is that it provided us with one of the oldest
counterexamples to the cosmic censorship conjecture. Indeed, in the
original paper of Papapetrou \cite{Papapetrou}, it was shown that
this solution could give rise to the formation of naked
singularities. In addition, since  its discovery, the Vaidya
solution has been intensively studied and also generalized in
presence of source. In particular its electrical extension \cite{BV}
has given rise to numerous researches whether from a thermodynamic
point of view \cite{SI} or about the possible violation of the weak
energy condition \cite{Ori}. Extension of Vaidya solutions in the
case of Lovelock gravity and their thermodynamics features were also
discussed in \cite{Cai:2008mh}.

In the present article, we are interested in searching for Vaidya
type solutions within the framework of scalar tensor theories. An
example of such solution was exhibited  in the case of a particular
Horndeski theory in Ref. \cite{Babichev:2022awg}. As is now well
known, the Horndeski theory refers to the most general scalar tensor
theory leading to second-order equations for the metric and the
scalar field \cite{Horndeski:1974wa}. In the last decade this theory
has been "rediscovered", and a considerable progress has been made
in studying black holes in Horndeski theory. Recently, in Ref.
\cite{Babichev:2023dhs}, the authors have selected some sub-classes
of Horndeski theories by requiring that a certain combination of the
Einstein equations within a spherical ansatz splits into two
distinct branches. This splitting allows in some cases to determine
the form of the scalar field without knowing the metric function
explicitly. In this case, some integrability and compatibility
conditions have been established which permit to derive interesting
black hole solutions whose asymptotic behaviors resemble to those of
Schwarzschild-(A)dS. It is important for our work to mention that
these solutions depend on a single integration constant identified
with the mass of the black hole. The resulting action permitting
such solutions, is devoid of any apparent symmetry (such as the
shift symmetry), but surprisingly the scalar field
action\footnote{In what follows, by scalar field action, we mean the
terms in the action that depend on the scalar field in opposition
with those of  pure gravity.} can be seen as the sum of an action
yielding a conformally invariant scalar field equation (with
subscribes $4$) with a another piece that is is conformally
invariant in five dimensions (with subscribes $5$), {\footnotesize
\begin{eqnarray}
\label{eq:action}
S=&&\int d^4 x \sqrt{-g}\Biggl[R - 2\Lambda-2 \lambda_4 \mathrm{e}^{4 \phi}-2 \lambda_5 \mathrm{e}^{5 \phi}\\
&&-\beta_4 \mathrm{e}^{2 \phi}\left(R+6(\nabla \phi)^2\right)-\beta_5 \mathrm{e}^{3 \phi}\left(R+12(\nabla \phi)^2\right)\nonumber\\
&&-\alpha_4\left(\phi \mathcal{G}-4 G^{\mu \nu} \nabla_\mu\phi\nabla_\nu\phi-4 \square \phi(\nabla \phi)^2-2(\nabla \phi)^4\right)\nonumber\\
&&-\alpha_5 \mathrm{e}^\phi\left(\mathcal{G}-8 G^{\mu \nu}
\nabla_\mu\phi\nabla_\nu\phi-12 \square \phi(\nabla
\phi)^2-12(\nabla \phi)^4\right)\Biggr]\nonumber.
\end{eqnarray}
} Note that for the vanishing parameters with subscribes $5$, this
action reduces to the one first considerer in
\cite{Fernandes:2021dsb}, and for which two classes of black hole
solutions have been found.

First, we will generalize the integrability conditions given in
\cite{Babichev:2023dhs} for a spherical metric (written in
Eddington-Finkelstein) whose base manifold has constant curvature
$\kappa=\pm 1$ or $\kappa=0$, namely
\begin{eqnarray}
\label{ansatzmetric} d s^2=-f(r)\,du^2 - 2\,du\,dr +\frac{r^2
d\theta^2}{1-\kappa \theta^2}+r^2 \theta^2 d y^2.
\end{eqnarray}
The topological black holes generalizing the solutions found in
\cite{Babichev:2023dhs} will be exhibited. A particular attention is
devoted to the planar case $\kappa=0$ where we  show that the
integrability conditions impose that the couplings
$\lambda_4=\lambda_5=\beta_4=\beta_5=0$. In this planar case, we
will also see that the scalar field has a free parameter
unconstrained by the field equations. The reason of this freedom
will be clarified. In a second time, we show that demanding the
metric function $f$ to depend as well on the retarded (advanced)
time $u$, the generalization of the integrability conditions
naturally lead to Vaidya-like metric. More precisely, we establish
that each topological black holes can be promoted to Vaidya-like
metric by promoting the constant mass to a function of $u$. In the
third part, we extend our analysis in higher dimensions where the
action is now given by the Lovelock Lagrangian with a scalar field
action that is conformally invariant. We first see that a certain
combinations of the equations of motion within our ansatz of the
form (\ref{ansatzmetric}) splits in two branches as in the Einstein
gravity case. Similarly, we establish that the Lovelock static black
hole solutions of these theories found in \cite{Giribet:2014bva} and
\cite{Babichev:2023rhn}, can as well be promoted to Vaidya-like
solutions by allowing the mass to depend on the coordinate $u$.

%%%%%%%%%%%%%%%%%%%%%%%%%%%%%%%%%%%%%%%%%%%%%%%%%%%%%%%%%%%%%%%%
\section{Integrability conditions for topological black holes}
%%%%%%%%%%%%%%%%%%%%%%%%%%%%%%%%%%%%%%%%%%%%%%%%%%%%%%%%%%%%%%%

Let us denote by
\begin{eqnarray}
E_{\mu\nu}:=G_{\mu\nu}+\Lambda g_{\mu\nu}-T_{\mu\nu},
\end{eqnarray}
the field equations arising from the variation of the action
(\ref{eq:action}) with respect to the metric, and whose explicit
expressions are reported in the Appendix. For an ansatz metric of
the form (\ref{ansatzmetric}) with a scalar field depending only on
the radial coordinate $r$, the equation $E_{rr}=0$ splits in two
radically distinct branches given by
\begin{equation}
\begin{split}
{}&{}\left(\phi'^2 - \phi''\right)\biggl[r^2\left(2 \beta_4+3 \beta_5 e^{\phi}\right) e^{2\phi}\\
{}&{}+4\alpha_5\left(\kappa -f\left(1+r \phi^{\prime}\right)\left(1+3 r \phi^{\prime}\right)\right) e^{\phi}\\
{}&{}+4\alpha_4\left(\kappa-f\left(1+r\phi^{\prime}\right)^2
\right)\biggr]=0\label{eq:master}.
\end{split}
\end{equation}
The first branch will refer to the equation $\left(\phi'^2 -
\phi''\right)=0$ for which the integration of the scalar field can
be done without knowing {\it a priori} the expression of the metric
function. For the second branch, as already mentioned in
\cite{Babichev:2023dhs}, the integration of the scalar field can
only be achieved in the case $\alpha_5=\beta_5=\lambda_5=0$, and the
factorization (\ref{eq:master}) reduces to
\begin{equation}
\begin{split}
{}&{}\left(\phi'^2 - \phi''\right)\biggl[\beta_4 r^2
e^{2\phi}+2\alpha_4\left(\kappa-f\left(1+r\phi^{\prime}\right)^2
\right)\biggr]=0\label{eq:master2}.
\end{split}
\end{equation}

In what follows, we  first consider the first branch for which we
give the compatibility conditions and exhibit as well its
topological black holes. In the second sub-section, we will
establish that  these static topological black hole solutions can be
extended to Vaidya-like solutions by turning the constant mass to be
an arbitrary function of the retarded time $u$. Although the second
branch is somewhat different in that the scalar field depends
explicitly on the constant $M$ and the couplings with the subscribe
$5$ are taken to zero, we will see that a similar analysis will
surprisingly lead to the same conclusions.

%%%%%%%%%%%%%%%%%%%%%%%%%%%%%%%%%%%%%%%%%%%%%%%%%%%%%%%%%%%%%%%%%%%
\subsection{ First branch of topological black hole solutions}
%%%%%%%%%%%%%%%%%%%%%%%%%%%%%%%%%%%%%%%%%%%%%%%%%%%%%%%%%%%%%%%%%%%%%
Here, we focus on the solution given by the first branch of
(\ref{eq:master}), namely $\phi'^2 - \phi''=0$. Although the general
solution should include two integration constants, the full
integration of the system will impose one of them to be zero and,
hence we anticipate the following form for the scalar field solution
\begin{equation}
\phi(r)=\ln{\left(\frac{\eta}{r}\right)}, \label{eq: primera g}
\end{equation}
where $\eta$ is {\it a priori} a free integration of constant.
Plugging this ansatz into the remaining Einstein equations, $E_{uu}$
and $E_{\theta\theta}$, one obtains that
\begin{align}
\label{Estatic}
{E^{\mbox{{\tiny{static}}}}_{uu}}&=\Biggl[\frac{\alpha_4 f^2}{r}-\left(r+\frac{\beta_5 \eta^3}{2 r^2}+\frac{2 \kappa \alpha_5 \eta}{r^2}+\frac{2 \alpha_4 \kappa}{r}\right) f\nonumber\\
 &+\frac{1}{r^2}\left(\frac{\lambda_5 \eta^5}{2}+\frac{\beta_5 \eta^3 \kappa}{2}\right)+\frac{1}{r}\left(\lambda_4 \eta^4+\beta_4 \kappa \eta^2\right)\nonumber\\
 &-\frac{r^3 \Lambda}{3}+\kappa r\Biggr]^{\prime},\\
{E^{\mbox{{\tiny{static}}}}_{\theta\theta}}&=\Biggl[\frac{\alpha_4}{r} f^2-\left(r-\frac{\beta_5 \eta^3}{r^2}-\frac{\beta_4 \eta^2}{r}\right) f-\frac{\lambda_5 \eta^5}{3 r^2}\nonumber\\
&+\frac{1}{r}\left(-\lambda_4 \eta^4\right)-\frac{r^3
\Lambda}{3}\Biggr]^{\prime \prime}.\nonumber
\end{align}
Integrating both equations we obtain two quadratic  equations for
the metric $f$,
\begin{subequations}
\label{Estatic2}
\begin{eqnarray}
&&\frac{\alpha_4 f^2}{r}-\left(r+\frac{\beta_5 \eta^3}{2 r^2}+\frac{2 \kappa \alpha_5 \eta}{r^2}+\frac{2 \alpha_4 \kappa}{r}\right) f\nonumber\\
&&+\frac{1}{r^2}\left(\frac{\lambda_5 \eta^5}{2}+\frac{\beta_5 \eta^3 \kappa}{2}\right)+\frac{1}{r}\left(\lambda_4 \eta^4+\beta_4 \kappa \eta^2 \right)\nonumber\\
&&-\frac{r^3
\Lambda}{3}+\kappa r + C_1=0,\\
\nonumber\\
&&\frac{\alpha_4}{r} f^2-\left(r-\frac{\beta_5 \eta^3}{r^2}-\frac{\beta_4 \eta^2}{r}\right) f-\frac{\lambda_5 \eta^5}{3 r^2}\nonumber\\
&&-\frac{\lambda_4 \eta^4}{r}-\frac{r^3 \Lambda}{3}+C_3r + C_2=0,
\end{eqnarray}
\end{subequations}
where $C_1$, $C_2$ and $C_3$ are a priori three different
integration constants. In order for these equations to be
compatibles, the coupling constants of the system are forced to be
tied as follows
\begin{eqnarray}
\label{cccon} \beta_5 =-\frac{4}{3 \eta^2}\alpha_5\kappa,
\nonumber\,\,\,\,\,\,\beta_4   = - \frac{2\alpha_4}{\eta^2} \kappa,\\
\lambda_5=-\frac{3}{5 \eta^2}\beta_5\kappa, \,\,\,\,\,\,\lambda_4
=-\frac{1}{2\eta^2}\beta_4\kappa,\\ \nonumber
\end{eqnarray}
while the constants of integration must be fixed as
\begin{eqnarray}
\label{ccconint} C_3=\kappa,\,\,\,\,\,\, C_1=C_2=-2M,
\end{eqnarray}
and, where $M$ is a truly integration constant. Note that in the
planar case $\kappa=0$, the couplings $\beta_4, \lambda_4, \beta_5$
and $\lambda_5$ must vanish (\ref{cccon}), and this case will be
treated separately below. Otherwise for $\kappa=\pm 1$, the constant
$\eta$ of the scalar field becomes fixed in term of the coupling
constants of the theory, and defining
\begin{eqnarray}
\label{intestatic} {\cal E}^{\mbox{{\tiny{static}}}}=&&
\frac{\alpha_4
f^2}{r}-\left(r+\frac{4\alpha_5\eta\kappa}{3r^2}+\frac{2\alpha_4\kappa}{r}\right)f-\frac{4 \eta}{15r^2}\alpha_5\kappa^2\nonumber\\
&&-\frac{\alpha_4\kappa^2}{r}-\frac{r^3\Lambda}{3}+\kappa r,
\end{eqnarray}
the two quadratic equations for the metric function (\ref{Estatic2})
at the "point" defined by the compatibility conditions
(\ref{cccon}-\ref{ccconint}) become a single relation
$$
{\cal E}^{\mbox{{\tiny{static}}}}-2M=0,
$$
whose general metric solution can be parameterized  in terms of
$\alpha_4\not=0$, $\alpha_5$ and $\eta$ as
\begin{equation}
\label{sol1}
\begin{split}
  f(r)&=\kappa+\frac{2 \alpha_5 \eta \kappa}{3 r \alpha_4}+\frac{r^2}{2\alpha_4}\Biggl(1\pm \Biggr[\left(1+\frac{4 \alpha_5 \eta \kappa}{3 r^3}\right)^2\\
  &+4 \alpha_4\Biggl(\frac{\Lambda}{3}+\frac{2M}{r^3}+\frac{2 \alpha_4 \kappa^2}{r^4}+\frac{8  \alpha_5 \eta \kappa^2}{5 r^5}\Biggr)\Biggr]^{\frac{1}{2}}\Biggr),
\end{split}
\end{equation}
On the other hand, for $\alpha_4=0$, the equation defining $f$
becomes linear and in this case, the topological black hole metric
function is given by
\begin{align}
\label{sol2} f(r)&=\frac{1}{1+\frac{4 \alpha_5 \eta\kappa}{3
r^3}}\left[\kappa-\frac{\Lambda r^2}{3}-\frac{2 M}{r}-\frac{4
\alpha_5 \eta\kappa^2}{15 r^3}\right].
\end{align}
Note that both solutions (\ref{sol1}-\ref{sol2}) correspond to those
reported in \cite{Babichev:2023dhs} for $\kappa=1$.

 Let us now go back in more details to the planar case
$\kappa=0$. As said before, in this case the couplings
$\beta_4=\lambda_4=\beta_5=\lambda_5=0$, and
%the compatible theory is then given by {\small
%\begin{equation}
%    \begin{split}
%        S{}&{}=\int d^4 x \sqrt{-g}\Biggl\{R - 2\Lambda-\alpha_4\Biggl(\phi \mathcal{G}-4 G^{\mu \nu} \phi_\mu \phi_\nu\\
%        {}&{}-4 \square \phi(\nabla \phi)^2-2(\nabla \phi)^4\Biggr)-\alpha_5 e^\phi\Biggl(\mathcal{G}-8 G^{\mu \nu} \phi_\mu \phi_\nu\\
%        {}&{}-12 \square \phi(\nabla \phi)^2
%        -12(\nabla \phi)^4\Biggr)\Biggr\}.
%    \end{split}
%\end{equation}
%}
consequently the field equations reduce to
\begin{eqnarray}
\label{eqko} &&G_{\mu\nu}+\Lambda g_{\mu\nu}=\alpha_4
\mathcal{H}_{\mu
\nu}^{(4)}+\alpha_5 e^{\phi} \mathcal{H}_{\mu \nu}^{(5)},\\
&& \alpha_4 \mathcal{H}^{(4)}+\alpha_5
\mathrm{e}^{\phi}\mathcal{H}^{(5)}=0,\nonumber
\end{eqnarray}
where the different expressions of the tensors can be found in the
Appendix. It is a matter of check to see that the following metric
function and scalar field
\begin{equation}
\label{sol3}
\begin{split}
  f(r)&=\frac{r^2}{2\alpha_4}\Biggl(1\pm \sqrt{ 1+4 \alpha_4\left(\frac{\Lambda}{3}+\frac{2M}{r^3}\right)}\Biggr),\\
 \phi(r)&=\ln\left(\frac{\tilde{\eta}}{r}\right),
\end{split}
\end{equation}
satisfy the equations (\ref{eqko}) in the planar case $\kappa=0$.
Various comments can be made concerning this solution. First, one
can see that although the coupling $\alpha_5\not=0$, it does not
appear  in the expression of the metric neither in that of the
scalar field. Also, in contrast with the cases $\kappa=\pm1$, the
constant $\tilde{\eta}$ is a truly integration constant
unconstrained by the field equations. These two features of the
solution (\ref{sol3}) can be explained by the fact that the
configuration (\ref{sol3}), in addition to satisfying the equations
of motion (\ref{eqko}), makes also that the stress tensor
$\mathcal{H}_{\mu \nu}^{(5)}$ and the scalar quantity
$\mathcal{H}^{(5)}$ to identically vanish on-shell. Consequently,
this explains the absence of the coupling $\alpha_5$ in the
parametrization of the solution. Concerning the constant
$\tilde{\eta}$, it is clear that, since the $\alpha_4-$part of the
action is invariant under the shift constant of the scalar field
$\phi\to \phi+\mbox{cst}$, and since the $\alpha_5-$part of the
equations vanish on-shell, our scalar field solution will always be
defined up to a constant.

%%%%%%%%%%%%%%%%%%%%%%%%%%%%%%%%%%%%%%%%%%%%%%%%%%%%%%%%%%%
\subsection{The Vaidya-like extension of the first branch of solutions}
%%%%%%%%%%%%%%%%%%%%%%%%%%%%%%%%%%%%%%%%%%%%%%%%%%%%%%%%%%%%
We now show that all the previous solutions can be extended to
Vaidya type solutions thanks to a generalization of the previous
comparability relations. In order to achieve this task, we allow the
metric function $f$ to depend as well on the retarded time but we
restrict the form of the scalar field as in the static
case\footnote{One could also have considered a scalar field of the
form $\phi=\phi(u,r)$, and this will considerably complicate the
problem, but for our purpose it is enough to consider
$\phi=\phi(r)$.}
\begin{eqnarray}
\label{Vaidyaansatz}
&&ds^2=-f(u, r) du^2-2dudr+\frac{r^2d\theta^2}{1-\kappa \theta^2}+r^2 \theta^2d y^2,\nonumber\\
&&\phi(r)=\ln\left(\frac{\eta}{r}\right),
\end{eqnarray}
where $\eta$ will be fixed  by the compatibility conditions  as in
the static case. As a consequence of this choice, the Einstein
equation ${E_{rr}}=0$ is automatically satisfied since the first
part of the factorization (\ref{eq:master}) is unchanged. Now,
evaluating the expression of the scalar field into the remaining
independent Einstein equations at the special tuning point where the
topological black holes exist (\ref{cccon}), one gets
\begin{eqnarray}
\label{EVaidya}
&&{E_{uu}}=\frac{1}{r^2}\left(f(u,r)\partial_r-\partial_u\right){\cal
E}^{\mbox{{\tiny{static}}}}{(f(u,r))},
\nonumber\\
 &&E_{\theta
\theta}=\frac{r}{2(\kappa \theta^2-1)}\partial_{rr}{\cal
E}^{\mbox{{\tiny{static}}}}(f(u,r)),
\end{eqnarray}
where  ${\cal E}^{\mbox{{\tiny{static}}}}(f(u,r))$ refers to the
expression defined in (\ref{intestatic}) but now evaluated at
$f=f(u,r)$. It is clear that this system is incompatible unless
${\cal E}^{\mbox{{\tiny{static}}}}(f(u,r))$ is a constant, and in
this case one would end up with the static solution previously
derived. Nevertheless, let us explore other possibilities, and for
this we first focus on the equation $E_{\theta \theta}=0$. Its
general solution is given by ${\cal
E}^{\mbox{{\tiny{static}}}}(f(u,r))=C_1(u)r+2M(u)$ where $C_1(u)$
and $M(u)$ are two arbitrary functions of $u$. Injecting this
expression into the equation ${E_{uu}}$, one gets
\begin{eqnarray}
\label{EuuV}
&&{E_{uu}}=\frac{f(u,r)\,C_1(u)}{r^2}-\frac{\dot{C}_1(u)}{r}-\frac{2\dot{M}(u)}{r^2}.
\end{eqnarray}
A straightforward computation shows that the compatibility of the
equation $E_{uu}=0$ together with the expression (\ref{intestatic})
would yield the previous static solution, namely $C_1(u)=0$ and
$M(u)=M=\mbox{cst}$. However, one can opt for the option that this
configuration behaves as a pure Vaidya-like radiation field. Indeed,
this can occur by choosing $C_1(u)=0$ and leaving free the
dependence of the function $M(u)$. Indeed, in this case, the full
Einstein equations become
\begin{eqnarray}
\label{vvl} E_{\mu\nu}:=G_{\mu\nu}+\Lambda
g_{\mu\nu}-T_{\mu\nu}=-\frac{2\dot{M}(u)}{r^2}\delta_{\mu}^u\delta_{\nu}^u,
\end{eqnarray}
for an ansatz of the form (\ref{Vaidyaansatz}) with a metric
function given by
\begin{eqnarray}
\label{solvv}
&&f(u,r)=\kappa+\frac{2 \alpha_5 \eta \kappa}{3 r \alpha_4}+\frac{r^2}{2\alpha_4}\Biggl(1\pm \Biggr[\left(1+\frac{4 \alpha_5 \eta \kappa}{3 r^3}\right)^2\nonumber\\
&&+4 \alpha_4\Biggl(\frac{\Lambda}{3}+\frac{2M(u)}{r^3}+\frac{2
\alpha_4 \kappa^2}{r^4}+\frac{8  \alpha_5 \eta \kappa^2}{5
  r^5}\Biggr)\Biggr]^{\frac{1}{2}}\Biggr).
\end{eqnarray}
In sum, we have shown that the topological black hole $\kappa=\pm 1$
defined at the special point (\ref{cccon}) can naturally be promoted
to a Vaidya-like solution (\ref{vvl}) by promoting the constant mass
of the metric function to an arbitrary function of the retarded time
(\ref{solvv}).

Along the same lines, the $\kappa=0$ black hole solution
(\ref{sol2}) with the couplings $\alpha_4=\beta_4=\lambda_4=0$ can
as well rendered to satisfy the same Vaidya conditions (\ref{vvl})
for a metric function, and a scalar field given by
\begin{eqnarray*}
&&f(u,r)=\frac{1}{1+\frac{4 \alpha_5 \eta\kappa}{3
r^3}}\left[\kappa-\frac{\Lambda r^2}{3}-\frac{2 M(u)}{r}-\frac{4
\alpha_5 \eta\kappa^2}{15 r^3}\right],\\
&&\phi(r)=\ln\Big(\frac{\eta}{r}\Big).
\end{eqnarray*}
We also mention that for $M=M(u)$, the Vaidya extension of the
static solution with a planar base manifold  (\ref{sol3}) also
satisfy the stealth equations $\mathcal{H}_{\mu
\nu}^{(5)}=0=\mathcal{H}^{(5)}$ as in the static case.

%%%%%%%%%%%%%%%%%%%%%%%%%%%%%%%%%%%%%%%%
\subsection{Second branch of solutions}
%%%%%%%%%%%%%%%%%%%%%%%%%%%%%%%%%%%%%%%%
We now consider the second branch of equation (\ref{eq:master}).
Unfortunately, as shown in Ref. \cite{Babichev:2023dhs}, this branch
can only be solved analytically for $\alpha_5=\beta_5=\lambda_5=0$,
and in this case, the solution for $\kappa=1$ was given in
\cite{Fernandes:2021dsb}. Its topological generalization  can be
conveniently parameterized as follows
\begin{eqnarray}
\label{topo2}
&&f(r)=\kappa+\frac{r^2}{2\alpha_4}\left[1\pm \sqrt{1+4\alpha_4\left(\frac{2M}{r^3}+\frac{\Lambda}{3}\right)}\right],\\
\nonumber\\
&&\phi(r)=\ln\Bigg(\frac{\sqrt{-\frac{{2\kappa}\alpha_4}{\beta_4}}+(1-\kappa^2)\sqrt{\frac{{2\kappa}\alpha_4}{\beta_4}}}
{r\sinh\left[\sqrt{\kappa}\left(c_1\pm\int^r\frac{dr}{r\sqrt{f(r)}}\right)\right]}\Bigg),\nonumber
\end{eqnarray}
and the solution holds at the special fine tuning
$\lambda_4=3\beta_4^2/4\alpha_4$. Note that for this second branch,
the emergence of an unconstrained constant of integration $c_1$, a
sort of hair. For this second branch, it is clear that the scalar
field depends on the mass parameter $M$ through the expression of
the metric function $f$. Moreover, in contrast with the previous
case, if one would naively turn on the time dependence of the metric
function and on the scalar by promoting the mass $M$ to a function
$M(u)$, the Einstein field equation $E_{rr}=0$ would not be
satisfied. This is mainly because for a time-dependent ansatz for
the scalar field, $\phi(u,r)$, the static equation (\ref{eq:master})
for $\alpha_5=\beta_5=\lambda_5=0$ becomes factorized as {\footnotesize 
\begin{eqnarray}
\label{secondb}
&&\left[(\partial_r\phi)^2-\partial_{rr}\phi\right]\times\\
&&\left[\beta_4 r^2 e^{2\phi} + 2\alpha_4 \{\kappa-(1+r\partial_r\phi)(f(1+r\partial_r\phi)-2r\partial_u
\phi)\}\right]=0,\nonumber
\end{eqnarray}}
and, because of the presence of the term $\partial_u \phi$, the
second branch will not be satisfied by just turning the constant $M$
to a function $M(u)$. Despite this inconvenient, it can \pagebreak
be shown
that the Einstein's equations for a non-static ansatz of the form
(\ref{Vaidyaansatz}) with a metric function
$$
f(u, r)=\kappa+\frac{r^2}{2\alpha_4}\left[1\pm
\sqrt{1+4\alpha_4\left(\frac{2M(u)}{r^3}+\frac{\Lambda}{3}\right)}\right]
$$
are such that
\begin{eqnarray}
\label{vvl2nd} E_{\mu\nu}:=G_{\mu\nu}+\Lambda
g_{\mu\nu}-T_{\mu\nu}=-\frac{2\dot{M}(u)}{r^2}\delta_{\mu}^u\delta_{\nu}^u,
\end{eqnarray}
provided that the scalar field satisfied the equation of the second
branch (\ref{secondb}), namely{\footnotesize
\begin{eqnarray}
\label{ssfeq} \beta_4 r^2 e^{2\phi} + 2\alpha_4 \left[\kappa-(1+r\partial_r\phi)(f(1+r\partial_r\phi)-2r\partial_u
\phi)\right]=0.
\end{eqnarray}}
Note that the spherical case $\kappa=1$ was already reported along
the same lines in \cite{Babichev:2022awg}.

%%%%%%%%%%%%%%%%%%%%%%%%%%%%%%%%%%%%%%%%%%
\section{Extension to higher dimensions}
%%%%%%%%%%%%%%%%%%%%%%%%%%%%%%%%%%%%%%%%%

As already mentioned, the scalar field equation of the action
(\ref{eq:action}) for $\alpha_5=\beta_5=\lambda_5=0$, enjoys the
conformal invariance although the scalar field action is not
conformally invariant \cite{Fernandes:2021dsb}. Recently, this kind
of symmetry present at the level of the equation and not at the
level of the action has been dubbed non-Noetherian conformal
symmetry \cite{Ayon-Beato:2023bzp}. In Ref. \cite{Babichev:2022awg}, the authors
presented a procedure to obtain this non-Noetherian conformally
invariant action from a Noetherian conformally action in higher
dimensions. This latter action, whose pure gravity action included
the Lovelock theory, was first considered in \cite{Oliva:2011np} and
\cite{Giribet:2014bva}, and is nothing but the most general theory
of gravity conformally coupled to a scalar field that yields
second-order field equations for the scalar field and the metric. In
order to be self-contained, we will present the action with the
useful notations as introduced in Refs. \cite{Oliva:2011np} and
\cite{Giribet:2014bva}.

Let us first define the following tensor
\begin{eqnarray*}
{S_{\mu \nu}}^{\gamma \delta}&&=\Phi^2 {R_{\mu \nu}}^{\gamma
\delta}-4 \Phi \delta_{[\mu}^{[\gamma} \nabla_{\nu]}
\nabla^{\delta]} \Phi+8 \delta_{[\mu}^{[\gamma} \nabla_{\nu]} \phi
\nabla^{\delta]} \Phi \\
&&- 2 \delta_{[\mu}^{[\gamma} \delta_{\nu]}^{\delta]} \nabla_\rho
\Phi \nabla^\rho \Phi.
\end{eqnarray*}
where now the scalar field $\Phi$ is related to the previous one
$\phi$ by $\Phi=e^{\phi}$, and the action under consideration is
given by{\small
\begin{equation}
S=\int d^D x
\sqrt{-g}\Biggl\{\sum_{k=0}^{\left[\frac{D-1}{2}\right]}
\frac{1}{2^k} \delta^{(k)}\left(a_k R^{(k)}+b_k \Phi^{D-4 k}
S^{(k)}\right)
     \Biggr\},\label{Actiong}
\end{equation}}
where the  $a_k$ and $b_k$ are coupling constants. The Kronecker
tensor $\delta^{(k)}$ is defined by
\begin{equation*}
\delta^{(k)}=k ! \delta_{\left[\alpha_1\right.}^{\mu_1}
\delta_{\beta_1}^{\nu_1} \ldots \delta_{\alpha_k}^{\mu_k}
\delta_{\left.\beta_k\right]}^{\nu_k},
\end{equation*}
while the expressions of $R^{(k)}$ and $S^{(k)}$ read
\begin{equation*}
R^{(k)}=\prod_{r=1}^k{R_{\mu_r \nu_r}}^{\alpha_r \beta_r}, \quad
S^{(k)}=\prod_{r=1}^k {S_{\mu_r \nu_r}}^{\alpha_r \beta_r}.
\end{equation*}
As in four dimensions, the Einstein equation $E_{rr}=0$ for a static
ansatz of the form (\ref{ansatzmetric}) with a radial scalar field
can be factorized in a similar form of (\ref{eq:master}). On the
other hand, static black hole solutions similar to those described
previously were constructed in Refs. \cite{Giribet:2014bva,
Babichev:2023rhn}. In \cite{Babichev:2023rhn}, it was shown that
solutions of the action (\ref{Actiong}) can be projected through a
limiting process to the four-dimensional solutions of the action
(\ref{eq:action}) with $\alpha_5=\beta_5=\lambda_5=0$. It is
therefore natural to ask whether the static high-dimensional
solutions can be converted into Vaidya-type solutions which would
project into those previously derived. This is indeed the case, and
these Vaidya-like solutions for the first branch read
\begin{eqnarray}
\label{VaidyaansatzD}
&&ds^2=-f(u, r) du^2-2dudr+r^2d\Omega^2_{D-2, \kappa},\nonumber\\
&&\Phi(r)=\frac{\eta}{r},
\end{eqnarray}
where the metric function $f(u, r)$ satisfies a polynomial equation
of order $\left[ \frac{D-1}{2}\right]$ that reads
\begin{align}
&\sum\limits_{k=0}^{\left[  \frac{D-1}{2}\right] }\frac{a_k(D-1)!}
{(D-2k-1)!}\left(  \frac{\kappa-f(u,r)
}{r^{2}}\right)^{k} =\nonumber\\
&\frac{M(u)(D-1)(D-2)}{ r^{D-1}} -\frac{q(D-1)(D-2)}{r^D}.
\label{poly}
\end{align}
Here, the constant $\eta$ is defined in term of the coupling
constants of the Lagrangian through the relation
\begin{align}
\sum\limits_{k=1}^{\left[  \frac{D-1}{2}\right]  }k\
\frac{{b}_{k}}{(D-2k-1)!} \kappa^{k-1}\eta^{2-2k}  &
=0\text{,}\label{ConstN}
\end{align}
and the constant $q$ is given by
\begin{eqnarray}
q=-\frac{b_0^{(i)}}{(D-2)}\eta^D-\sum\limits_{k=1}^{\left[
\frac{D-1}{2}\right] }\frac{b_k(D-3)!\kappa^k}{(D-2k-2)!}
\eta^{D-2k}. \label{qi}
\end{eqnarray}
It is a matter of check to see that injecting the solutions given by
(\ref{VaidyaansatzD}-\ref{qi}) into the field equations of the
action (\ref{Actiong}) will give rise the Vaidya conditions
\begin{eqnarray}
\label{vaiidyad} {\cal G}_{\mu\nu}-T_{\mu\nu}=-\frac{(D-2)
\dot{M}(u)}{r^{D-2}}\delta_{\mu}^u\delta_{\nu}^u,
\end{eqnarray}
where ${\cal G}_{\mu\nu}$ is the Lovelock tensor.  As anticipated,
the metric solution of the polynomial equation (\ref{poly}) by means
of the limiting process $D\to 4$ described in
\cite{Babichev:2023rhn} will yield to (\ref{solvv}) with the
couplings $\alpha_5=\beta_5=\lambda_5=0$.

%%%%%%%%%%%%%%%%%%%%%%%
\section{Conclusions}
%%%%%%%%%%%%%%%%%%%%%%
In general, it is rather  difficult to find time-dependent
spherically symmetric solutions to Einstein's equations in the
presence of a some matter source. There are, however, a few
examples, such as stealth configurations, see e. g.
\cite{Ayon-Beato:2004nzi, Babichev:2013cya, Hassaine:2013cma} and
\cite{Babichev:2022awg},  or even the example of time-dependent
spherically symmetric solution that describes the gravitational
collapse to a scalar black hole in three dimensions, see
\cite{Xu:2014xqa} and \cite{Ayon-Beato:2015ada}. Here, we have
considered some classes of scalar tensor theories such that a
certain combination of the Einstein equations can be factorized out
as (\ref{eq:master}) within a spherical ansatz of the form
(\ref{ansatzmetric}). For these these theories, we have shown that
from static black hole configurations, and  by extending their mass
parameter to a function of the retarded time, one can end-up with
Vaidya-like configurations satisfying
$$
{\cal G}_{\mu\nu}-T_{\mu\nu}=-\frac{(D-2)
\dot{M}(u)}{r^{D-2}}\delta_{\mu}^u\delta_{\nu}^u.
$$
In general, it is a nontrivial task to find  matter that may source
the Vaidya geometries, that is some source that compensates  the
right-hand side of the previous equation by means of its
energy-momentum tensor. For example, as shown in
\cite{Faraoni:2021zin},  it would be impossible for a massless
scalar field minimally coupled to Einstein gravity, and this even if
the scalar field  with lightlike gradient behaves like a pure
radiation field.

On the other hand, we are convinced that promoting the mass constant
to an arbitrary function of the retarded time will not always yield
to Vaidya-like configurations. In fact, the possibility of
generating such Vaidya-type solutions from static solutions in our
case is essentially due to the factorization of the equation as
given in (\ref{eq:master2}) together with the fact that the scalar
field solution of the first branch does not depend on $M$. In order
to reinforce our intuition, we can consider the examples of the
static black hole solution of a conformally scalar field known as
the BBMB solution \cite{BBM, Bekenstein:1974sf} or its
self-interacting extension \cite{Martinez:2002ru}. These both
theories are particular cases of those considered here since they
correspond  to the action (\ref{Actiong}) with
$\lambda_5=\beta_5=\alpha_4=\alpha_5=\lambda_4=0$ (and in the
self-interacting case $\lambda_4\not=0$). Nevertheless, the main
differences are due to the fact since $\alpha_4=0$, the
factorization (\ref{eq:master2}) yields only to the first branch,
and, in this case, the static scalar field solution depends
explicitly on the mass constant $M$. It is then a matter of check to
see that even by promoting the constant $M$ to an arbitrary function
of time $M=M(u)$ the full equations will yield inconsistencies
unless $M=\mbox{cst}$.

%It therefore seems necessary to try to understand in a deeper way
%the reason why the static solutions that we have selected here could
%be extended to Vaidya type configurations.

%%%%%%%%%%
\appendix
%%%%%%%%%%

%%%%%%%%%%%%%%%%%%%%%%%%%%%%%%%%%
\section{Field equations}
%%%%%%%%%%%%%%%%%%%%%%%%%%%%%%

The field equations obtained by varying the action (\ref{eq:action})
with respect to the metric read
\begin{eqnarray}
G_{\mu\nu}+ \Lambda g_{\mu\nu}=  T_{\mu\nu} \label{fieldeqs}
\end{eqnarray}
where
\begin{eqnarray*}
&&T_{\mu\nu}=\beta_4 \mathrm{e}^{2\phi}\mathcal{A}_{\mu\nu}^{(4)} +
\alpha_4\mathcal{H}_{\mu\nu}^{(4)}-\lambda_4
\mathrm{e}^{4\phi}g_{\mu\nu}+ \beta_5
\mathrm{e}^{3\phi}\mathcal{A}_{\mu\nu}^{(5)} \nonumber\\
&&+ \alpha_5
\mathrm{e}^\phi\mathcal{H}_{\mu\nu}^{(5)}-\lambda_5\mathrm{e}^{5\phi}g_{\mu\nu}
\end{eqnarray*}
where the terms $\mathcal{A}_{\mu\nu}^{(4)}$,
$\mathcal{A}_{\mu\nu}^{(5)}$, $\mathcal{H}_{\mu\nu}^{(4)}$ and
$\mathcal{H}_{\mu\nu}^{(5)}$ are those associated with the
energy-momentum tensor of the scalar field, and are given by {\small
\begin{equation}
\begin{split}
\mathcal{A}_{\mu \nu}^{(4)}{}&{}=G_{\mu \nu}+2 \nabla_\mu \phi
\nabla_\nu \phi-2 \nabla_\mu \nabla_\nu \phi
+g_{\mu \nu}\left(2 \square \phi+(\nabla \phi)^2\right),\\
\mathcal{A}_{\mu\nu}^{(5)} {}&{}
=G_{\mu\nu}+3\nabla_{\mu}\phi\nabla_{\nu}\phi-3\nabla_{\mu}\nabla_{\nu}\phi+g_{\mu
\nu}\left(3\square \phi+(\nabla \phi)^2\right)\nonumber,
    \end{split}
\end{equation}}
{\small
\begin{equation}
\label{st}
\begin{split}
\mathcal{H}_{\mu \nu}^{(4)}{}&{}= -2G_{\mu \nu}(\nabla \phi)^2+4 P_{\mu \alpha \nu \beta}\left(\nabla^\alpha \nabla^\beta \phi-\nabla^\alpha \phi \nabla^\beta \phi\right)\\
{}&{}+4\left(\nabla_\alpha \phi \nabla_\mu \phi-\nabla_\alpha \nabla_\mu \phi\right)\left(\nabla^\alpha \phi \nabla_\nu \phi-\nabla^\alpha \nabla_\nu \phi\right) \\
{}&{} +4\left(\nabla_\mu \phi \nabla_\nu \phi-\nabla_\nu \nabla_\mu \phi\right) \square \phi+g_{\mu \nu}\left(-2(\square \phi)^2+(\nabla \phi)^4\right)\\
{}&{}+2g_{\mu \nu}\nabla_\beta \nabla_\alpha \phi\left(\nabla^\beta \nabla^\alpha \phi-2 \nabla^\alpha \phi \nabla^\beta \phi\right),\\
 \mathcal{H}_{\mu \nu}^{(5)}{}&{}=-4G_{\mu \nu}(\nabla \phi)^2+4 P_{\mu \alpha \nu \beta}\left(\nabla^\alpha \nabla^\beta \phi-\nabla^\alpha \phi \nabla^\beta \phi\right)\\
 {}&{}+ 8\left(\nabla_\mu \phi \nabla_\nu \phi-\nabla_\nu \nabla_\mu \phi\right) \square \phi-4g_{\mu \nu}\square \phi\left(\square \phi+(\nabla \phi)^2\right)\\
{}&{}+4g_{\mu \nu}\nabla_\alpha \nabla_\beta \phi\left(\nabla^\alpha \nabla^\beta \phi-2 \nabla^\alpha \phi \nabla^\beta \phi\right)\\
{}&{}+8\left(\nabla_{\mu}\phi\nabla_{\nu}\nabla_{\alpha}\phi\nabla^{\alpha}\phi+\nabla_{\nu}\phi\nabla_{\mu}\nabla_{\alpha}\phi\nabla^{\alpha}\phi\right)\\
{}&{}-8\nabla_{\mu}\nabla_{\alpha}\phi\nabla_{\nu}\nabla^{\alpha}\phi+4\left(\nabla\phi\right)^2\left(\nabla_{\mu}\nabla_{\nu}\phi-3\nabla_{\mu}\phi\nabla_{\nu}\phi\right).
\end{split}
\end{equation}}\\
Here, the tensor $P_{\mu\alpha\nu\beta}$ is defined as follows{\small
\begin{equation}
\begin{split}
P_{\mu\alpha\nu\beta}{}&{}= R_{\mu\alpha\nu\beta}+g_{\mu\beta}R_{\alpha\nu}+g_{\alpha\nu}R_{\mu\beta}-g_{\mu\nu}R_{\alpha\beta}-g_{\alpha\beta}R_{\mu\nu}\\
{}&{}+\frac{1}{2}\left(g_{\mu\nu}g_{\alpha\beta}-g_{\mu\beta}g_{\alpha\nu}\right)R.\nonumber\label{eq:p}
\end{split}
\end{equation}}
The variation of the action (\ref{eq:action}) with respect to the
scalar field yields
\begin{align}
 0{}&{}=\beta_4\mathrm{e}^{2\phi}\mathcal{A}^{(4)} + \alpha_4 \mathrm{e}^{\phi}\mathcal{H}^{(4)}+8\lambda_4\mathrm{e}^{4\phi}+\beta_5\mathrm{e}^{3\phi}\mathcal{A}^{(5)}\nonumber\\
{}&{}+ \alpha_5 \mathrm{e}^{\phi}\mathcal{H}^{(5)}+10\lambda_5
\mathrm{e}^{5\phi},\label{eq:scalar_eq}
\end{align}
where $\mathcal{A}^{(4)}$, $\mathcal{A}^{(5)}$, $\mathcal{H}^{(4)}$
and  $\mathcal{H}^{(5)}$ are is given by
\begin{equation}
    \begin{split}
    \mathcal{A}^{(4)}{}&{}= 2\left(R-6\Box\phi-6\left(\nabla\phi\right)^2\right),\\
    \mathcal{A}^{(5)}{}&{}= 3\left(R-8\Box\phi-12\left(\nabla\phi\right)^2\right)\nonumber,
    \end{split}
\end{equation}
and where  {\small
\begin{equation}
    \begin{split}
    \mathcal{H}^{(4)}{}&{} =\mathcal{G}+8\left(G^{\mu\nu}\nabla_{\mu}\nabla_{\nu}\phi-R^{\mu\nu}\nabla_{\mu}\phi\nabla_{\nu}\phi + \Box\phi\left(\nabla\phi\right)^2\right)\\
    {}&{}+8\left(2\nabla^{\mu}\phi\nabla_{\mu}\nabla_{\nu}\phi\nabla^{\nu}-\nabla_{\mu}\nabla_{\nu}\phi\nabla^{\mu}\nabla^{\nu}\phi+\phi(\Box\phi)^2\right),\\
    \mathcal{H}^{(5)}{}&{} = \mathcal{G}+16\left(G^{\mu\nu}\nabla_{\mu}\nabla_{\nu}\phi-R^{\mu\nu}\nabla_{\mu}\phi\nabla_{\nu}\phi + 3\Box\phi\left(\nabla\phi\right)^2\right)\\
    {}&{}+24\left(\left(\left(\Box\phi\right)^2-\nabla_{\mu}\nabla_{\nu}\phi\nabla^{\mu}\nabla^{\nu}\phi+\left(\nabla\phi\right)^4\right)\right)\\
    {}&{}-4 R\left(\nabla\phi\right)^2 +48\nabla^{\mu}\phi\nabla_{\mu}\nabla_{\nu}\phi\nabla^{\nu}\phi.\nonumber
\end{split}
\end{equation}}

%The external matter stress tensor needed to support this solution is

%%%%%%%%%%%%%%%%%%%%%%%%%%%%%%%%%%%%%%%%%%%%%%%%%%%%%%%%%%%%%%%%%%%%
\begin{acknowledgments}

We would like to thank Eloy Ay\'on-Beato and Julio Oliva for
interesting discussions. This work has been partially funded by
FONDECYT grant $1210889$ and ANID grant $21231297$.
\end{acknowledgments}
%%%%%%%%%%%%%%%%%%%%%%%%%%%%%%%%%%%%%%%%%%%%%%%%%%%%%%%%%%%%%%%%%%%%

%%%%%%%%%%%%%%%%%%%%%%%%%%%%%%

\end{document}